\documentclass[11pt]{article}
\usepackage{amsmath,epsf,amssymb,latexsym,enumerate,cite,shadow,array,color}
\usepackage{slashed}
\usepackage{graphicx,wasysym}

\DeclareFontFamily{U}{rsf}{} \DeclareFontShape{U}{rsf}{m}{n}{
  <5> <6> rsfs5 <7> <8> <9> rsfs7 <10-> rsfs10}{}
\DeclareMathAlphabet\Scr{U}{rsf}{m}{n} \makeatletter
\@addtoreset{equation}{section} \makeatother

\def\be{\begin{equation}}
\def\ee{\end{equation}}
\def\ba{\begin{array}}
\def\ea{\end{array}}

\newcommand{\bea}{\begin{eqnarray}}
\newcommand{\eea}{\end{eqnarray}}

\def\K{K{\"a}hler}
\usepackage{ifpdf}
\ifx\pdfoutput\undefined
   \pdffalse
\else
   \pdfoutput=1
   \pdftrue
  \usepackage[pdftex]{hyperref}
\def\u0{{\underline 0}}

\def\url{{\underline {r+\ell}}}

\newcommand{\rf}[1]{(\ref{#1})}

\begin{document}

\begin{titlepage}

\hskip 1cm

\vskip 3cm

\begin{center}
{\LARGE \textbf{Inflation and Uplifting with Nilpotent Superfields\\
\vskip 0.8cm }}

\

  {\bf Renata Kallosh}\,   {\bf and  Andrei Linde}
\vskip 0.5cm
{\small\sl\noindent  Department of Physics, Stanford University, Stanford, CA
94305 USA}
\end{center}
\vskip 0.5 cm

\

\begin{abstract}
Recently it was found that a broad class of existing inflationary models based on supergravity can be significantly simplified if some of the standard, unconstrained chiral superfields are replaced by nilpotent superfields, associated with Volkov-Akulov supersymmetry. The same method allows to simplify the existing models of uplifting of AdS vacua in string theory. In this paper we will show that one can go well beyond simplifying the models that already exist. We will propose a broad class of new models of chaotic inflation  based on supergravity with nilpotent superfields, which simultaneously incorporate both inflation and uplifting. They provide a simple unified description of inflation and the present acceleration of the universe in the supergravity context. 
\end{abstract}

\vspace{24pt}
\end{titlepage}

\newpage

\section{Introduction}

The simplest model capable of describing both inflation and the present stage of acceleration of the universe has the potential
\be\label{infuplift}
V = {m^{2}\over 2}\phi^{2} +V_{0} \ .
\ee
In the early universe, the first term dominates, as in the simplest chaotic inflation scenario \cite{Linde:1983gd}. Then the field $\phi$ decays and disappears, density of matter produced after inflation gradually decreases, and the energy density becomes dominated by the tiny cosmological constant $V_{0}\sim 10^{-120}$, which leads to the present stage of the accelerated expansion of the universe. More generally, one could also consider chaotic inflation with potentials
\be\label{moregen}
V = f^{2}(\phi) + V_{0}\ , 
\ee
where $f(\phi)$ is some function vanishing in its minimum. However, for a long time realistic supergravity-based inflationary models with such potentials were unavailable.

A significant progress in this respect was achieved when a simple supergravity realization of inflation with the quadratic potential ${m^{2}\over 2}\phi^{2}$ was proposed \cite{Kawasaki:2000yn}. This scenario was substantially generalized in \cite{Kallosh:2010xz} to include nearly arbitrary chaotic inflation potentials. One may, for example, consider models with the following \K\, potential and superpotential:
\be\label{model1}
K= K[(\Phi-\bar\Phi)^2,S\bar S] \ , \qquad  W = S f(\Phi) \  .
\ee
Here $S$ and $\Phi$ are chiral superfields, and $f(\Phi)$ is an arbitrary real holomorphic function. The field $S$ is sometimes called the stabilizer. Representing the scalar component of the superfield $\Phi$ as  $(\phi+ i\, a)/\sqrt 2$, and the scalar component of $S$ as $\sigma\, e^{i\theta}/\sqrt 2$, one finds that under certain conditions  the fields $\sigma$ and $a$ vanish during inflation, and the field $\phi$ plays the role of the inflaton field with the potential \cite{Kallosh:2010xz}
\be
V(\phi) = |f(\phi/\sqrt 2)|^{2}\ .
\ee
Stabilization of the fields $\sigma$ and $a$ during inflation can be achieved by a proper choice of the \K\, potential, which does not affect the shape of the inflaton potential $V(\phi)$. For a broad class of models, the field $a$ has a large mass during inflation, it rapidly rolls down to the minimum of its potential at $a = 0$ and disappears. However, in many inflationary models the field $S$ requires stabilization, which can be achieved e.g. by adding a sufficiently large term $\sim (S\bar S)^{2}$ to the \K\, potential. Adding higher order terms to the \K\, potential is not an unreasonable price to pay for the functional freedom of the inflationary potential. However, it would be nice to avoid this problem altogether.

In this paper we will discuss the possibility to get rid of the stabilized field(s) by making them a part of the nilpotent multiplet, as proposed  in \cite{Ferrara:2014kva} in application to a broad class of existing models of inflation in supergravity. The idea,  which was first first outlined in \cite{Antoniadis:2014oya} in the context of the Starobinsky model, is to use the Volkov-Akulov mechanism of a nonlinear realization of supersymmetry \cite{Volkov:1973ix}, which allows to have supersymmetry without fundamental scalars. This means that instead of the usual unconstrained chiral multiplets $S(x, \theta)$ one should take nilpotent chiral multiplets with $S^{2}(x, \theta) = 0$, \cite{rocek},\cite{Komargodski:2009rz}. For a detailed formulation of this approach in supergravity, we refer the readers to \cite{Ferrara:2014kva} and the references therein. Here we will simply formulate the main rule of constructing a bosonic part of the supergravity action in this approach. The derivation of this rule is not trivial, but the final result is extremely simple:
One should take the theory of several chiral multiplets, define their \K\, potential and superpotential, calculate the scalar potential, kinetic terms, etc. as one usually does. When all computations are completed, one should declare that the vev of the scalar field belonging to the nilpotent chiral multiplet vanishes \cite{Ferrara:2014kva}.

In application to the inflationary models which we study, one may, for example, consider a nilpotent stabilizer multiplet $S$, calculate everything, and then, instead of investigating time evolution and stability of its scalar component $s$, one should simply declare that its vev vanishes. Of course, there is a price for this miracle: The field does not really disappear, it becomes replaced by a bilinear combination of fermions, but its expectation value vanishes. More work is needed in order to correctly describe the fermionic part of the action,  but for the description of  inflation one may concentrate on the bosonic part of the theory.  

For a very large class of models mentioned in \cite{Ferrara:2014kva}, the upshot is very simple: Inflationary models with unconstrained chiral superfields stabilized at $S(x, \theta)|_{\theta=0}=s(x) = 0$,  will continue working in exactly the same way if $S$ is a nilpotent superfield, but without any need to stabilize the scalar part of the superfield $S$: Its vev vanishes by construction.

The same method was applied in \cite{Ferrara:2014kva} to the theory of uplifting in string theory landscape. One of the approaches to the F-term uplifting was to add to the theory describing the KKLT construction \cite{Kachru:2003aw} a Polonyi-type field with a linear superpotential, stabilize this field, and use it for uplifting of the minimum of the potential from AdS to dS,   see e.g. \cite{Lebedev:2006qq,Kallosh:2006dv,Dudas:2012wi,Kallosh:2014oja}.  Now the same can be done much easier: After the uplifting, the scalar component of the nilpotent Polonyi field disappears, no need to stabilize it. This also means that the Polonyi field participating in the uplifting does not lead to the famous cosmological Polonyi field problem. Moreover, as emphasized in  \cite{Ferrara:2014kva}, this mechanism has an interesting interpretation relating the Volkov-Akulov approach to physics of Dp-branes. The fermions which live on the world-volume of the D-branes, in general,  are the Volkov-Akulov goldstinos, which at the  supergravity level can be described by the nilpotent chiral multiplets.

In the recent study of the  KKLT model uplifting in \cite{Kallosh:2014wsa}, without inflation, it was  realized  that the supersymmetric uplifting requires orientifolding, equivalent to a supersymmetric truncation.  This new analysis is consistent with the fact that D3-brane is not responsible for the uplifting.  It is the presence of the anti-D3-brane that leads to KKLT de Sitter vacua with Volkov-Akulov goldstino and spontaneous breaking of supersymmetry,
which brings us back to the supergravity with nilpotent superfields.

The remaining step is to apply the new approach to inflation and to uplifting together, using either the standard  superfields as well as the nilpotent superfields. In this paper we will develop simple supergravity models with nilpotent superfields, which describe both inflation and the present stage of the exponential acceleration of the universe.

\section{Inflation and uplifting with one nilpotent multiplet}

One could expect that adding a tiny constant $V_{0} \sim 10^{-120}$ to an inflationary potential should be very easy, given the functional freedom of choice of $f(\Phi)$ in \cite{Kallosh:2010xz}. Paradoxically, the smallness of $V_{0}$ makes this problem nontrivial. Consider, for example, the model of two standard, unconstrained chiral fields,
\be
K=-{(\Phi- \bar\Phi)^2\over 2} + S\bar S  , \qquad 
 W=  Sf(\Phi) \ ,
\label{1i}\ee
with $f(\Phi) = M^{2}(1+c \Phi^{2})$. According to \cite{Kallosh:2010xz}, the potential of the real part of the scalar field is $M^{4} (1+c \phi^{2}/2)^{2}$. 
At small $\phi$, this potential looks as $M^{4} + c M^{4} \phi^{2}$, which is very similar to \rf{infuplift}, so one could think that the problem is nearly solved.
However, one can show that for $c > 1$ the theory is unstable with respect to generation of an imaginary part of the field $\Phi$. The value of the potential in its true minimum vanishes, so it does not describe the accelerated universe. One can avoid this problem if $c< 1$, but then for $V_{0} = M^{4} \sim 10^{{-120}}$ one has an almost exactly flat inflationary potential, and the perturbations of metric produced in this scenario become vanishingly small. 

Of course it is possible to achieve uplifting by stabilizing the field $S$ and adding other superfields \cite{Lebedev:2006qq,Kallosh:2014oja},  but this makes the theory more complicated. Can we do it in a nice and easy way without introducing additional scalar fields and worrying about their time evolution and the cosmological moduli problems they may cause?

\subsection{Quadratic inflation and uplifting with $f(0) \not =0$}

The answer to the  question above is positive, if the superfield $S$ is nilpotent. Let us add a constant $W_{0}$ to the superpotential, 
\be
K=-{(\Phi- \bar\Phi)^2\over 2} + S\bar S  , \qquad 
 W=  SM^{2}(1+c \Phi^{2}) +W_{0}\ .
\label{1ii}\ee
 Because of the additional term $W_{0}$, this theory deviates from the structure of the models of  \cite{Kallosh:2010xz} where we had $W=  Sf(\Phi)$. If the field $S$ is unconstrained, it acquires a non-zero vev depending on $W_{0}$, and the potential $V$ in its absolute minimum becomes negative. However, if the superfield $S$ is nilpotent, its scalar component disappears by construction, so one should only study the potential of the remaining fields $\phi$ and $a$:
\be
V(\phi,a) ={e^{\,a^{2}}}  \Bigl[M^{4}(1+c\, (\phi^{2}-a^{2})) +W_{0}^{2}\, (2a^{2}-3)+{M^{4}c^{2}\over 4}(\phi^{2}+a^{2})^{2}\Bigr]\ .
\label{pot}\ee
The point $a = 0$ is always an extremum of the potential. We will assume that during inflation $a = 0$, and we will later find the conditions under which this assumption is valid. For $a = 0$, the inflaton potential becomes 
\be\label{quad}
V(\phi) =cM^{4} \phi^{2}\ \Bigl(1 +{c\over 4}\phi^{2}\Bigr) +V_{0} \ ,
\ee
where its value in the vacuum is given by
\be\label{cancel}
V_{0} =M^{4} -3W_{0}^{2}  \sim 10^{{-120}}\ .
\ee
We will assume that each of these two terms, $M^{4}$ and  $3W_{0}^{2}$, is much greater than the tiny cosmological constant $V_{0} \sim 10^{{-120}}$, but their cancelation allows to fine-tune the value of the vacuum energy in the same spirit as it is supposed to happen in the string theory landscape: String theory may provide us with an abundant choice of different vacua  \cite{Bousso:2000xa,Kachru:2003aw}. In some of them, different contributions to the vacuum energy almost exactly cancel, which makes the corresponding parts of the universe suitable for life.
The existence of the tiny cosmological constant does not affect the process of inflation, but in the course of the subsequent expansion of the universe, the energy density becomes dominated by the tiny cosmological constant $V_{0}$, which leads to the present stage of the nearly exponential expansion of the universe. 

In the theory with a quadratic potential, the last 60 e-folds of inflation occur for $\phi \lesssim 15$. Therefore for $c \ll 10^{-2}$ the quartic term in \rf{quad} is subdominant at that stage, and inflation occurs just like in the theory with the potential ${m^{2}\over 2}\phi^{2}$ with $m^{2} = {2c} M^{4}$. Thus the last 60 e-folds of inflation and the present stage of acceleration of the universe in this theory will be described by the model \rf{infuplift}. To match the COBE/Planck normalization for the amplitude of inflationary perturbations, one should have $m \sim 6 \times 10^{{-6}}$, which yields $c\, M^{4} \sim 2 \times 10^{-11}$. As an example satisfying all our constraints, one may take $c \sim 10^{{-4}}$ and $M \sim 0.02$.

The last point to check is the stability of the field $a$ at $a = 0$. The calculations of the mass squared of the field $a$ during inflation in this scenario for $c \ll 1$ shows that it is greater than $6H^{2}$, so the state $a =0$ is stable. The state $a = 0$ remains stable after inflation as well; the mass squared of the field $a$ at the minimum of the potential at $\phi = a = 0$ is given by ${4\over 3} M^{4}$.

The superpotential in the vacuum is given by $W_{0} = M^{2}/\sqrt 3 = 2.5\times 10^{-4}$. The function $DW$ has only one non-vanishing component  in the goldstino direction: $D_{S}W = M^{2}$.  The function $f(\Phi) =M^{2}(1+c \Phi^{2})$ in the minimum is also given by $M^{2}$. Note that this function does not vanish in the minimum. This is important from the point of view of the discussion of fermions in this model \cite{Ferrara:2014kva}.

This model at the minimum actually has a simple description of the  fermion part since $D_\Phi W=0$, gravitino  interacts with goldstino $\psi_S$ and does not interact with inflatino $\psi_\Phi$. This allows to make a choice of the local supersymmetry gauge $\psi_S=0$. All complicated non-linear terms depending on $\psi_S$ vanish in this gauge and at the minimum we recover the super-Higgs effect \cite{cfgvp,SUGRA}. Gravitino eats goldstino, becomes fat, its mass is $m_{3/2}^2= W_0^2$. The remaining inflatino has a mass $m_{1/2}^2= W_0^2$. The masses of the scalar  fields are  $m_{a}^2\approx  4 W_0^2$,  assuming that  $3 W_{0} ^2\approx  M^{4}$ and that $c\ll 1$, and $m_\phi^2=6 \, c \, W_0^2$.

\subsection{General class of inflationary models with uplifting with $f(0) \not =0$}

Let us now consider a more general scenario with $f(\Phi) = M^{2}(1+g^{2}(\Phi))$:
\be
K=-{(\Phi- \bar\Phi)^2\over 2} + S\bar S  , \qquad 
 W=  SM^{2}(1+g^{2}(\Phi)) +W_{0}\ .
\label{1}\ee
Here $g(\Phi)$ is a holomorphic function vanishing in the minimum of the potential at $\Phi =0$. For each such function, one should check the validity of the assumption that $a = 0$ is a stable solution of equations of motion before and after inflation, similarly to what we did above. But once it is done, the inflaton potential acquires the form
\be
V(\phi) =M^{4}|g(\phi/\sqrt 2)|^{2}\Bigl(2+|g(\phi/\sqrt 2)|^{2}\Bigr) +V_{0} \ ,
\ee
where, as before, $V_{0} =M^{4} -3W_{0}^{2}  \sim 10^{{-120}}$.
For $|g(\phi/\sqrt 2)|^{2}\ll 1$, the potential during inflation can be approximated by $V(\phi) =2M^{4}|g(\phi/\sqrt 2)|^{2}$. The values of $W$, $DW$, and $f(\Phi) = M^{2}(1+g(\Phi)^{2})$ in the vacuum at $\Phi =0$ coincide with the corresponding values in the model \rf{1ii} studied above. 

 The models \rf{1ii} and \rf{1} considered above are pretty simple, and allow a straightforward calculation of fermionic masses. However,  this simplicity comes at a price: The tiny uplifting of the inflationary potential by $V_{0} \sim 10^{{-120}}$ in this scenario requires strong supersymmetry breaking, with the gravitino mass greater than the inflation mass. Models of such type are internally consistent, but if one wants to have the gravitino mass below $10^{2}$ TeV, as suggested by many  (as yet unconfirmed) phenomenological models, one may try to do something else.

\subsection{Inflation and uplifting with $f(\Phi) =m\Phi$}

Now we will investigate a different model, where the uplifting occurs not because of the non-zero value of the function $f$, but because of the separate term in the superpotential proportional to $\Phi$.
We will start with the model \rf{model1} with the simplest function $f(\Phi) = m  \Phi$, with two additional terms in the superpotential, $M^{2} \Phi$ and $W_{0}$, such that $W_{0}^{2}, M^{4} \ll m^{2}$. 
\be
K=-{(\Phi- \bar\Phi)^2\over 2} + S\bar S  , \qquad 
 W=  m S \Phi + M^{2} \Phi + W_{0} \ .
\label{1u}\ee
The resulting potential, for zero scalar component of the superfield $S$,  is
\begin{eqnarray}
V &=&{e^{\,a^{2}}\over 2}  \Bigl[{m^{2}}(\phi^{2}+a^{2}) + 2\bigl(\sqrt 2\, W_{0}\,M^{2}\, \phi\,+W_{0}^{2}\bigr) (2a^{2}-3)\nonumber \\ & +& {M^{4}}\bigl(2+a^{2}+2a^{4}-3\phi^{2}+2\phi^{2}a^{2}\bigr)\Bigr]\ .
\label{2}\end{eqnarray}
As before, the potential has an extremum at $a = 0$. Let us assume first (and confirm shortly) that during inflation $a =0$. Then the inflaton potential is given by
\be
V(\phi)  ={m^{2}\over 2}\phi^{2}+ {M^{4}\over 2}\bigl(2-{3}\phi^{2}\bigr)-3W_{0}^{2}-3\sqrt 2 W_{0} M^{2} \phi\  \ .
\label{potgold}\ee
During inflation, it is dominated by its first term, $V\approx {m^{2}\over 2}\phi^{2}$.
The mass squared of the field $a$ for $a = 0$ is 
\be
m^{2}_{a} = m^{2}(1+\phi^{2}) +  M^{4}(3-\phi^{2})-2W_{0}^{2} -2\sqrt 2 W_{0} M^{2} \phi \ .
\label{23}\ee
In this case $m^{2}_{a} \approx m^{2}(1+\phi^{2}) > 0$, and during inflation with $a = 0$ it is greater than $6H^{2}$. Thus indeed the field $a$ rapidly rolls down to $a = 0$ and stays there.
The potential has a minimum at 
\be
\phi = {3\sqrt 2 W_{0}M^{2}\over m^{2}-M^{4}} \approx {3\sqrt 2 W_{0}M^{2}\over m^{2}}.
\label{7}\ee
The function $f(\Phi) = m\Phi \approx {3 W_{0}M^{2}\over m}$ does not vanish at the minimum. 
The value of the potential at the minimum (the cosmological constant) is 
\be
V_{0}={m^{2} (M^{4}-3 W_{0}^{2})-M^{8}\over m^{2}-M^{4}} \approx M^{4}-3 W_{0}^{2} +...  \ .
\label{8}\ee
where the omitted terms are much smaller than $M^{4}$ and $W_{0}^{2}$ for $W_{0}^{2}, M^{4} \ll m^{2}$. If $M^{4} \gg V_{0} \sim 10^{{-120}}$,  one has
\be
M^{4}\approx  3 W_{0}^{2} \ ,
\label{9}\ee
up to small corrections $O(M^{8}/m^{2},\, V_{0})\ll M^{4}$. This yields
\be
|\phi | \approx {3\sqrt 6 W_{0}^{2}\over m^{2}} \ll 1.
\label{10}\ee
In the same approximation, the value of the superpotential in the minimum is $W_{0}$, and the covariant derivatives of the superpotential are $D_{S}W \approx {\sqrt 3 M^{4}\over m^{2}}$, $D_{\Phi} W \approx M^{2}$.

Thus, we were able to almost exactly reproduce the simplest inflationary theory \rf{infuplift} describing also the present stage of acceleration in the context of supergravity with only one dynamical scalar field $\Phi$. The stabilizer field $S$ does not have any scalar degrees of freedom associated with it, and therefore does not lead to the cosmological moduli problem. 

 The calculation of the fermion masses in this scenario is somewhat more complicated than in the models \rf{1ii} and \rf{1}; we will return to it in a separate publication. However, we expect that the supersymmetry breaking in this scenario, unlike in the models \rf{1ii} and \rf{1}, can be made arbitrarily small. Indeed, in the limit $M^{2}, W_{0} \to 0$, there is no difference between the structure of this theory and the well studied models \cite{Kallosh:2010xz} with the stabilized field $S =0$ and the supersymmetric Minkowski vacuum $\Phi = 0$. Supersymmetry breaking appears only because of the parameters $M^{2}$ and $W_{0}$, which can be taken incredibly small, all the way down to $M^{4} \sim 3 W_{0}^{2} \sim 10^{{-120}}$. By tuning these parameters, while preserving the relation $M^{4}\approx  3 W_{0}^{2}$, one should be able to obtain the model with various values of vacuum energy, including $V_{0} \sim 10^{{-120}}$, and with a small level of supersymmetry breaking controlled by the choice of the small parameter $W_{0}$.

\section{Inflation and uplifting with two nilpotent multiplets}

\subsection{A general set of models}

As we are going to see now, the structure of the bosonic part of the inflationary theory becomes even simpler if we introduce two nilpotent superfields, $S$ and $T$, satisfying the conditions $S^2(x, \theta)= T^2(x, \theta)=0$. We will consider a theory 
\be
K=K[(\Phi-\bar\Phi)^2, S\bar S,T\bar T] \ , \qquad 
 W=  S f(\Phi) + M^{2} T + W_{0} \ .
\label{twogold}\ee
Just as in \cite{Kallosh:2010xz}, $f(\Phi)$ is a real holomorphic function, and we will assume that the imaginary part of the scalar component of the superfield $\Phi$ vanishes during inflation, $a = 0$. The stability of this field is to be confirmed by explicit calculations for each particular \K\ potential. In most of the cases presently studied in the literature, this field in the original theory (prior to adding the terms $M^{2}$ and $W_{0}$) does not require stabilization  \cite{Kallosh:2010xz}; this conclusion remains valid if the terms $M^{2}$ and $W_{0}$ are small enough. The scalar components of the nilpotent fields $S$ and $T$ vanish by construction. Without any loss of generality, one can always normalize the \K\, potential so that ${\cal K}(0,0,0,0) = 0$,\ ${{\cal K}_{S \bar S}}(0,0,0,0) = {{\cal K}_{\Phi \bar \Phi}}(0,0,0,0) =  1$, see  \cite{Kallosh:2010xz}.  The inflaton potential for $a =0$, $S|_{\theta=0}=s=0$ and $  T|_{\theta=0} =t= 0$ is 
\be
V =|f(\phi/\sqrt 2)|^{2} + M^{4}-3W_{0}^{2}\,\ .
\label{potgold5}\ee
If $f(\phi) = 0$ at the minimum of the potential, which is the case for the simplest functions $f(\Phi)$, then the vacuum energy $V(0) =|f(0)|^{2} + M^{4}-3W_{0}^{2}$ is given by
\be
V_{0} =M^{4}-3W_{0}^{2}\ .
\label{potgold2}\ee
The superpotential in the vacuum is $W_{0}$, and the only nonzero component of $DW$ is  $D_{T}W = M^{2}$. 

Note that the part $M^{2} T +W_{0}$ of the superpotential is very similar to the superpotential of the Polonyi field commonly used in supergravity phenomenology. However, unlike the Polonyi field, the nilpotent superfield $T$, just as the superfield $S$, does not have any scalar field component, and therefore these superfields do not lead to the cosmological moduli problems.

The level of supersymmetry breaking in this class of models, just like in the model \rf{1u}, is determined by the parameters $M^{2}$, and $W_{0}$, which can take any values such that $V_{0} =M^{4}-3W_{0}^{2} \sim 10^{{-120}}$. Thus, supersymmetry breaking in this class of models can be controllably small.

\subsection{The simplest example: Chaotic inflation with a quadratic potential and a cosmological constant}

The simplest example of models of the type of \rf{twogold} is chaotic inflation with
\be
K=-{(\Phi- \bar\Phi)^2\over 2} + S\bar S + T\bar T\, , \qquad 
 W=  m S \Phi + M^{2} T + W_{0} \ .
\label{twogolda}\ee
This model is very similar to the model \rf{1u} studied in the previous section. The main difference is that instead of the term $M^{2}\Phi$ in the superpotential, we have the term $M^{2} T$ involving the nilpotent field $T$. As we will see, this leads to considerable simplifications in our results.

The potential for vanishing scalar components of the superfields $S$ and $T$ is 
\be
V =e^{a^{2}} \left[{m^{2}\over 2}(\phi^{2}+a^{2}) + M^{4}+W_{0}^{2}\, (2a^{2}-3) \right]\ ,
\label{potgold5a}\ee
which is much simpler than the corresponding expression \rf{2}. The potential $V$ in this model at $a = 0$ coincides with the simplest potential \rf{infuplift} discussed in the Introduction:  
\be\label{quadr}
V ={m^{2}\over 2}\phi^{2} + V_{0}\ ,
\ee
with $V_{0} = M^{4}-3W_{0}^{2}$. The scalar field masses at $a = 0$ are
\be
m^{2}_{\phi} = m^{2} \,, \qquad
m^{2}_{a} = m^{2}(1+\phi^{2}) +  2M^{4}-2W_{0}^{2} \ .
\ee
Suppose $M^{4} \gg V_{0}\sim 10^{-120}$. Then one has $M^{4}=3W_{0}^{2}$ with accuracy $10^{-120}$, and therefore
$m^{2}_{a} = m^{2}(1+\phi^{2}) +  4W_{0}^{2} >0$.
During inflation one has $H^{2} = {V/3} = m^{2}\phi^{2}/6 + V_{0}/3 \approx m^{2}\phi^{2}/6$. Therefore
\be
m^{2}_{a} \approx 6H^{2} +m^{2}+4W_{0}^{2} > 6H^{2}\ ,
\ee
so the inflationary trajectory with $a = 0$ is strongly stabilized, and the only evolving field is, indeed, the inflaton field $\phi$. This confirms our main result, the expression for the inflaton potential \rf{quadr}.

\section{Discussion}

In this paper we concentrated on the bosonic sector of the new class of theories with nilpotent chiral superfields. For the full discussion of all related phenomenological models and evaluation of the possible limits of their applicability one should carefully investigate the fermionic sector.  This is an important issue, and we hope to return to it in the future.

The bosonic sector plays the dominant role in the discussion of inflation, uplifting, and the present state of a nearly exponential expansion. In this respect, the first results achieved so far are rather encouraging. The main idea is that one can introduce one or many different nilpotent supermultiplets without introducing any additional moduli. By using this possibility, one can simplify many of the existing inflationary models and construct new interesting cosmological models based on supergravity and string theory.

The inflationary models in \cite{Kawasaki:2000yn,Kallosh:2010xz,Ferrara:2014kva,Antoniadis:2014oya} based on the superpotential $W=S\, f(\Phi)$ typically had an  exit from inflation into a Minkowski vacuum with $f(\Phi)|_{\rm min}=0$ with $\Lambda=0$. In the new models developed in this paper the exit from  inflation is in  de Sitter space where at least one of the nilpotent superfields has a nonzero
 $DW=M^2$, and $W_0$ also does not vanish. In this case
\be
\Lambda = M^{4}-3W_{0}^{2} \ .
\ee
Here the positive energy $M^4$ is due to spontaneous supersymmetry breaking of Volkov-Akulov type originating from a nilpotent chiral multiplet, whereas $-3 W_0^2$ is the negative energy associated with spontaneous  supersymmetry breaking in the supergravity multiplet. The smallness of $\Lambda$ is due to an incomplete cancellation between these two fundamental contributions to supersymmetry breaking. 

This mechanism is in the spirit of the string theory landscape scenario \cite{Kachru:2003aw,Bousso:2000xa}.  However, now it is manifestly supersymmetric, with spontaneous breaking of supersymmetry, and the models also include generic inflation.

The new terms which we have now added to our models also have string theory interpretation: the $S$-independent terms in $W$ originate from fluxes and other string theory sectors not interacting with goldstinos. The additional nilpotent superfields  may be associated with goldstinos from additional D-branes.

The methods developed in this paper and  also in \cite{Ferrara:2014kva} revealed a broad class of new inflationary models, where the functional freedom of choice of the function $f(\Phi)$ allows one to account for any desirable values of the inflationary parameters $n_{s}$ and $r$   \cite{Kallosh:2010xz}, whereas the uplifting of the vacuum energy provides a consistent description of the present stage of the accelerated expansion of the universe. 
Theories of this class, as well as their generalizations with different \K\ potentials, such as $K=K[(\Phi+\bar\Phi)^2, S\bar S,T\bar T]$, or $K= -3\, \alpha \log \left(\Phi + \bar \Phi   -  S \bar S   \right)$, can describe uplifted versions of the theory of superconformal attractors, various supersymmetric generalizations of the Starobinsky model, supersymmetric realization of the Higgs inflation, natural inflation, and many other popular versions of inflationary cosmology.

Further investigation of D-brane physics in a consistent curved background with fluxes is required to construct string theoretical models  
related to d=4 N=1 supergravities with nilpotent multiplets \cite{Kallosh:2014wsa}. 
This is particularly important for developing new string theory inspired cosmological models compatible with the current and future observations.

\section*{Acknowledgments}

We are grateful to S. Ferrara, S. Kachru, P. Nilles, K. Olive, E. Silverstein and T. Wrase for many useful discussions. RK and AL are supported by the SITP and by the NSF Grant PHY-1316699 and RK is also supported by the Templeton foundation grant `Quantum Gravity Frontiers'.

\end{document}